\documentclass[onecolumn,showpacs,preprintnumbers,amsmath,amssymb,10pt]{revtex4}

\usepackage{epsf}
\usepackage{graphicx}  
\usepackage{dcolumn}   
\usepackage{bm}        

\newcommand{\be}{\begin{equation}}
\newcommand{\en}{\end{equation}}
\newcommand{\bea}{\begin{eqnarray}}
\newcommand{\ena}{\end{eqnarray}}
\newcommand{\lb}{\label}
\begin{document}

\preprint{SUCA/005-2010}

\title{De Sitter ground state of scalar-tensor gravity and its fluctuation with dust}

\author{Hongsheng Zhang\footnote{Electronic address: hongsheng@shnu.edu.cn} }
\affiliation{Shanghai United Center for Astrophysics (SUCA),
 Shanghai Normal University, 100 Guilin Road, Shanghai 200234,
 P.R.China}
  \author{Xin-Zhou Li \footnote{Electronic address: kychz@shnu.edu.cn} }
 \affiliation{Shanghai United Center for Astrophysics (SUCA), Shanghai Normal
University, 100 Guilin Road, Shanghai 200234, P.R.China}

\begin{abstract}
   An exact de Sitter solution of scalar-tensor gravity is found, in which the non-minimal
   coupling scalar is rolling along a non-constant potential. Based on this solution, a dust-filled FRW universe
   is explored in frame of scalar-tensor gravity. The effective dark energy induced by the sole non-minimal scalar
   can be quintessence-like, phantom-like, and more significantly,
   can cross the phantom divide.  The rich and varied properties of scalar-tensor gravity even with only one scalar is
   shown in this article.

\end{abstract}


\maketitle

\section{Introduction}

 The prediction of dark energy is one of the most significant cosmological breakthrough
 over the second half of last century \cite{acce}.  Dark energy is believed to be responsible for the present
 cosmic acceleration. Various models of dark energy have been
proposed, such as a small positive
 cosmological constant, quintessence, k-essence, phantom, holographic dark energy,
 etc., see \cite{review} for a
 recent review. However, although fundamental for
our understanding of the universe, its nature, especially in the
 theoretical aspect, remains a completely open question.

  Cosmological constant (CC) is a concordance and far simple candidate for dark energy. However, it suffers from
  two serious theoretical problems: fine-tuning problem and coincidence problem.
  Furthermore, more and more abundant observation data illuminate a remarkable possibility of dark energy: it may
  evolve in the history of the universe. Usually, we introduce a scalar field, called quintessence, to simulate the evolution of the dark
  energy, for details see the review article \cite{review}. A quintessence field is minimally coupled to gravity, ie, there is no
  product term of Ricci scalar $R$ and quintessence field $\phi$ in the  Lagrangian. The scalar-tensor gravity (STG) is one of the most influential competitors to
  Einstein¡¯s general relativity. Contrarily to common idea, it is compulsive in several cases that a non-minimal coupled scalar is involved
  in the propagation of gravity. The most competitive reason comes from quantum arguments. When quantum correction is considered, or renormalizability of the
  scalar field in curved space is required, the product term of $\phi^2 R$ is inevitable \cite{qcs}.  We always meet such terms when we make
  dimensional reduction from higher dimensional theory, such as Kaluza-Klein theory. The non-minimal coupling scalar has been
  investigated in cosmology, in context of inflation \cite{noninf} and in dark energy \cite{nondark}.

   The Minkowskian ground state is an obvious solution
  of STG. For any physical theories, to find exact mathematical solutions is an important topic. Next comes the physical interpretations of
  the solution thus obtained. The spherical symmetric solution of STG has been found in \cite{lilu}. Mathematically, de Sitter as the maximally space is
  undoubtedly important for any metric gravity theories. From observational side, recent studies illuminate that both the early universe (inflation) and the
  late-time universe (cosmic acceleration)  can be regarded as fluctuations on a de Sitter background. So de Sitter takes a pivotal status in gravity,
   especially in modern cosmology.  However, the de Sitter ground state of STG has not been found in literatures, though STG has been widely studied for several decades, and
    many ``almost de Sitter" cases have been proposed \cite{noninf}. We construct an exact de Sitter solution in STG, in which the scalar is rolling down from
  from a power-law potential. At the same time, the energy density and pressure are variable, which seems impossible for the minimal coupling case.
  This de Sitter ground state is important since both the early inflation and the late time acceleration can be regarded as fluctuations on
  a de Sitter background.

  This paper is organized as follows: In the next section we shall study the de Sitter solution for the STG.
  The dust fluctuation upon the de Sitter background, which is corresponding to the late time universe, is investigated in section  III.  We present our conclusion
     and some discussions in section IV.

  \section{de Sitter state for STG}

   We start from the action of STG,
   \be
   S= \int_{M} d^4 x\sqrt{-g} \left[
   {1\over 2\kappa}R-\frac{\xi}{2}\phi^2 R+L_{\rm scalar}(\phi)+L_{\rm matter}\right]+{1\over \kappa}\int_{\partial M} d^3 x\sqrt{-h} K.
       \label{actionSTG}
       \en
       Here $\kappa$ is the gravity constant, $M$ represents the spacetime manifold, $K$ denotes the extrinsic curvature of its boundary,
       $R$ is the Ricci scalar, $\xi$ is a constant, $h$, $g$ stand for the determinants of the 4-metric $g_{\mu\nu}$ and its induced 3-metric $h_{\mu\nu}$ on the boundary, respectively.
       $L_{\rm scalar}$ and $L_{\rm matter}$ are the Lagrangians of the non-minimally coupled scalar and other minimally coupled matters to gravity, respectively.
  $L_{\rm scalar}$ takes the same form of an ordinary scalar,
  \be
   L_{\rm scalar}=-\frac{1}{2}g^{\mu\nu}\nabla_{\mu}\phi\nabla_{\nu}\phi-V(\phi).
   \label{actionS}
   \en
   By making variation of (\ref{actionSTG}) with respect to $\phi$, one obtains the equation of motion of the scalar,
   \be
   g^{\mu\nu}\nabla_{\mu}\nabla_{\nu}\phi-\xi R\phi-\frac{dV}{d\phi}=0.
   \label{eom}
   \en
  To the equation of motion for $\phi$, the non-minimal coupling term  $\xi\phi^2R/2$ is just equivalent to an extra potential.
  Variation of (\ref{actionSTG}) with respect
  to $g_{\mu\nu}$ yields the field equation,

  \be
  (1-\kappa \xi \phi^2)G^{\mu\nu}=\kappa\left[T^{\mu\nu}(\phi)+T^{\mu\nu}({\rm matter})\right],
  \label{FE}
  \en
 where $G^{\mu\nu}$ denotes Einstein tensor, $T^{\mu\nu}({\rm matter})$ labels the energy-momentum tensor corresponding to
 $L_{\rm matter}$ in (\ref{actionSTG}), and $T^{\mu\nu}(\phi)$ takes the following form,

 \be
 T^{\mu\nu}(\phi)=\nabla^{\mu}\phi\nabla^{\nu}\phi-\frac{1}{2}g^{\mu\nu}g^{\alpha\beta}\nabla_{\alpha}\phi\nabla_{\beta}\phi-
 Vg^{\mu\nu}+\xi\left[g^{\mu\nu}g^{\alpha\beta}\nabla_{\alpha}\nabla_{\beta}(\phi^2)-\nabla^{\mu}\nabla^{\nu}(\phi^2)\right].
 \label{emnmt}
 \en
 For a detailed deduction of (\ref{FE}), see the appendix C in \cite{book}.  It deserves to note that one does not need to introduce new boundary
 term other than $K$ in this derivation, since one can repeatedly apply integration by parts to remove all the derivation terms of $g_{\mu\nu}$
 yielded by $\phi^2R$ on the boundary. If we define an effective gravity constant $\kappa_{\rm eff}$,
 \be
 \kappa_{\rm eff}=(1-\kappa \xi \phi^2)^{-1}\kappa,
 \en
 then the field equation (\ref{FE}) reduces to  Einstein form with a variable gravity ``constant". That is the original idea of Brans-Dicke proposal.
 When $\phi\to 0$, STG degenerates to standard general relativity. An hence both Minkowski and (anti-)de Sitter are permitted. But it is only a trivial case of
 STG. How about a non-zero $\phi$?

 Now we try to find the solution of STG with maximally symmetric space. A maximally space (in a proper chat) can be defined as an FRW universe with constant Hubble parameter.
 In an FRW universe, the field equation becomes Friedmann equations,
 \be
 (1-\kappa \xi \phi^2)\left(H^2+\frac{k}{a^2}\right)=\frac{\kappa}{3}\rho,
 \label{fried1}
 \en
 \be
 (1-\kappa \xi \phi^2)\frac{\ddot{a}}{a}=-\frac{\kappa}{6}(\rho+3p),
 \label{fried2}
 \en
 where $a$ is the scale factor, the factor $(1-\kappa \xi \phi^2)$ signifies that $\phi$ is involved in gravity interaction, $\rho$ and $p$ denote
 the total density and pressure,
 \be
 \rho=\frac{1}{2}\dot{\phi}^2+V+6\xi \phi \dot{\phi}H+\rho_{\rm matter},
 \en
 \be
 p=\frac{1-4\xi}{2}\dot{\phi}^2-V-2\xi \phi \ddot{\phi}-4\xi \phi \dot{\phi}H+p_{\rm matter}.
  \en
 Here matter labels all the matters other than $\phi$. To find the ground state of STG, we  consider the its vacuum solution without any other
 matter fields other $\phi$. In the next section we shall study a dust fluctuation on this background.  We derive the following solution of (\ref{fried1}) and
 (\ref{fried2}) by setting $\rho_{\rm matter}=0$ and $p_{\rm matter}=0$,
 \be
 \phi=c_2\left[-e^{2c_1b\xi}+e^{bt}(4\xi-1)\right]^{2\xi \over {4\xi-1}},
 \label{phis}
 \en
 \be
 V=\frac{3b^2}{\kappa}-\frac{\xi b^2}{\phi^2 (1-4\xi)^2}\left[(3-34\xi+96\xi^2)\phi^4+
 8\xi c_2^{2-\frac{1}{2\xi}}e^{2c_1b\xi}(6\xi-1)\phi^{2+\frac{1}{2\xi}}+2\xi c_2^{4-\frac{1}{\xi}}e^{4c_1b\xi}\phi^{\frac{1}{\xi}}\right],
 \label{potential}
 \en

 \be
 \rho=-p=\frac{3b^2}{\kappa}(1-\kappa \xi \phi^2),
 \label{rhop}
 \en

 \be
 a=c_3e^{bt},
 \en
  \be
  k=0,
  \en

 where $c_1,~c_2,~c_3$ are integration constants. It is easy to see that the above set-up describes a de Sitter space.
 Interestingly, though the density and pressure (\ref{rhop}) are not constant in the evolution of the spacetime, they keep
 cunning counteraction with the extra factor in the modified Friedmann equation (\ref{fried1}) and
 (\ref{fried2}). And hence the Hubble parameter can be a constant with a rolling scalar along the potential (\ref{potential}).
 The conformal coupling case is the most important case in the non-minimal coupling theory. For the conformal coupling case $\xi=1/6$,
  the potential $V$ degenerates to an extraordinary simple form,
  \be
  V= \frac{3b^2}{\kappa}-\frac{e^{2c_1b/3}b^2}{2c_2^2}\phi^4,
  \label{vsimple}
  \en
 while the other quantities in the above set-up become,
 \be
 \phi=-c_2\left(e^{c_1b/3}+\frac{1}{3}e^{bt}\right)^{-1},
 \en

 \be
 \rho=-p=\frac{3b^2}{\kappa}(1-\frac{1}{6}\kappa  \phi^2),
 \label{rhosim}
 \en

 \be
 a=c_3e^{bt},
 \en
  \be
  k=0.
  \en


 \section{dust fluctuation on the de Sitter background}
 De Sitter space becomes very important in modern cosmology, since both the inflation in the early
 universe and the cosmic acceleration in the late time universe are fluctuations on a de Sitter background. To study inflation or
 late-time acceleration require a different energy scale $b$ in (\ref{phis}). In the present article,
 we constrain ourself in the conformal coupling case and first study the late-time universe.

 Considering an FRW universe in STG filled with dust, one reaches,
  \be
 \rho=\frac{1}{2}\dot{\phi}^2+V+6\xi \phi \dot{\phi}H+\rho_{\rm dust},
 \en
 \be
 p=\frac{1-4\xi}{2}\dot{\phi}^2-V-2\xi \phi \ddot{\phi}-4\xi \phi \dot{\phi}H.
  \en
 From the field equation of STG (\ref{FE}), we derive
 \be
 R=\kappa (1-\kappa \xi \phi^2)^{-1}(\rho-3p).
 \en
 Through the investigations in Section I, we know that conformal coupling STG has an exact de Sitter
 phase under the potential (\ref{vsimple}).  Enlightened this investigation, we set the potential $V$ to study the current acceleration,
 \be
 \frac{\kappa}{3H_0^2}V=n-lx^4,
 \label{vnew}
 \en

 since our present universe can be treated as a perturbation of a de Sitter universe. In (\ref{vnew}), we set two dimensionless constant $n$ and $l$
 for further phenomenological explore.
 We take the Friedmann equation (\ref{fried1}) and the equation of motion of $\phi$ (\ref{eom}) as the fundamental set.
 In this set, there are two independent ordinary differential equations with two functions $a(t),~\phi{(t)}$ to solve.
 For convenience, we use $H(t)$ and $\phi(t)$ as fundamental equations to solve by making variable replacement. And then,
 we define two dimensionless variables $x,~y$,
 \be
 x=\sqrt{\kappa}\phi,
 \en
 \be
 y=\frac{H}{H_0},
 \en
 where $H_0$ stands for the current Hubble parameter. Equations (\ref{fried1}) and (\ref{eom}) reduce to
 \be
 (1-\frac{x^2}{6})y^2=\frac{1}{6}y^2x'^2+n-lx^4+\frac{1}{3}y^2xx'+\Omega_{m0}e^{-3s},
 \label{friedm}
 \en
 \be
 yy'x'+y^2x''+3y^2x'+\frac{1}{6}N_{t}-12lx^3=0.
 \label{eomm}
 \en
 Here $N_{t}$ corresponds to the non-minimal coupling term,
 \be
 N_{t}=x(1-\frac{1}{6}x^2)^{-1}(12n-12lx^4+3y^2xx'+xx'yy'+xy^2x''),
 \en
 $s\triangleq \ln a$, a prime denotes derivation with respect to $s$.
 It is difficult to find the analytical solution the set (\ref{friedm}) and (\ref{eomm}), we solve it by using numerical method.
 We shall see that even with a very simple potential (\ref{vnew}), the property of dark energy in
 this model is very rich and varied.

 First we present a concise note on the definition of dark energy.  In the STG theory,
   there is a surplus term $(1-\frac{1}{6}\kappa  \phi^2)$ in the modified Friedmann equation (\ref{fried1}).
    However, almost all observed properties of dark energy are
    obtained in frame of general relativity.
    To explain the the observed evolving  EOS of the effective dark
    energy,  we introduce the concept ``equivalent
 dark energy" or ``virtual dark energy" in the modified gravity
 models \cite{reviewcross}.  We derive the density of virtual dark energy caused by the non-minimal coupled scalar by comparing the modified Friedmann equation in
  the brane world scenario with the standard Friedmann equation in general
  relativity.
   The generic Friedmann equation in the
 4-dimensional general relativity can be written as
 \be
 H^2+\frac{k}{a^2}=\frac{\kappa}{3} (\rho_{dm}+\rho_{de}),
 \label{genericF}
 \en
 where the first term of RHS in the above equation represents the dust matter and the second
 term stands for the dark energy. Comparing (\ref{genericF})
 with (\ref{fried1}), one obtains the density of virtual dark
 energy in STG,
 \be
 \rho_{de}=(1-\frac{1}{6}\kappa  \phi^2)^{-1}\left(\frac{1}{2}\dot{\phi}^2+V+6\xi \phi \dot{\phi}H+\rho_{\rm matter}\right)-\rho_{\rm matter}.
 \label{rhode}
 \en
 Note that we consider the case that $\rho_{\rm matter}$ only comprises dust matter.
 For convenience, we introduce dimensionless dark energy,
 \be
 u=\frac{\kappa\rho_{de}}{3H_0^2}=(1-\frac{x^2}{6})^{-1}\left(\frac{1}{6}y^2x'^2+n-lx^4+\frac{1}{3}y^2xx'+\Omega_{m0}e^{-3s}\right)-\widetilde{\Omega}_{m0}e^{-3s}.
 \en
 The initial condition at present requires $H=H_0$, i.e.,
 \be
 1=y_0^2=(1-\frac{x_0^2}{6})^{-1}\left(\frac{1}{6}y_0^2x_0'^2+n-lx_0^4+\frac{1}{3}y_0^2x_0x_0'+\Omega_{m0}\right).
 \en
 We see that the unique reasonable $\widetilde{\Omega}_{m0}$ should be
 \be
  \widetilde{\Omega}_{m0}=(1-\frac{x_0^2}{6})^{-1}\Omega_{m0},
  \en
 if we require dark energy is completely yielded by $\phi$ at present epoch.

 Since the dust matter obeys the continuity equation
 and the Bianchi identity keeps valid, dark energy itself satisfies
  the continuity equation
 \be
 \frac{d\rho_{de}}{dt}+3H(\rho_{de}+p_{eff})=0,
 \label{contieff}
 \en
 where $p_{eff}$ denotes the effective pressure of the dark energy.
 And then we can express the equation of state for the dark
 energy as
   \be
  w_{de}=\frac{p_{eff}}{\rho_{de}}=-1-\frac{1}{3\rho_{de}}\frac{d \rho_{de}}{d \ln
  a}.
  \label{wde}
   \en
   From the above equation we find that the behavior of $w_{de}$ is determined by the term $\frac{d \rho_{de}}{d \ln
  a}$. $\frac{d \rho_{de}}{d \ln
  a}=0$ (cosmological constant) bounds phantom and quintessence. More
  intuitively, if $\rho_{de}$ increases with the expansion of the universe, the dark energy
  behave as phantom;  if $\rho_{de}$ decreases with the expansion of the universe, the dark energy
  behave as quintessence; if $\rho_{de}$ decreases and then increases, or increases and then
  decreases, we are certain that EOS of dark energy crosses phantom
  divide. A more important reason why we use the density to describe
  property of dark energy is that the density is
  more closely related to observables, hence is more tightly
  constrained for the same number of redshift bins used \cite{wangyun}.
 With data accumulation, observations which favor dynamical dark
   energy become more and accurate. Usually a quintessence, i.e., a canonical scalar field
   dominated by potential, can satisfy the observation. Furthermore, some data analysis
   implies that the present EOS of dark energy is less than $-1$. A phantom field, i.e., a scalar field with
   false kinetic term, can describe such an evolution. A significant possibility
   appears recently: the EOS of dark energy may cross $-1$ (phantom divide) \cite{vari}, which is a serious challenge for theoretical
   physics. The theoretical explore of the crossing phenomenon was proposed in \cite{cros}.
   This interesting topic is under intensively studying very recently \cite{cross}.

   We find that the effective dark energy (\ref{rhode}) can evolve as
 quintessence, phantom, or even cross the phantom divide. In the following text, we show our numerical results.
 In figs 1-3 we show the quintessence-like evolution of dark energy in STG. Fig 1 displays the evolution of the
 density of dark energy, fig 2 illuminate the corresponding EOS. Fig 3 displays the corresponding deceleration parameter $q$.
 The deceleration parameter is a most important parameter from the observation side, which carries the total effects of the
 fluids in the universe.

 In figs 4-6 we show the phantom-like evolution of dark energy. Fig 4 displays the evolution of the
 density of dark energy, figs 5 and 6 illuminate the corresponding EOS and deceleration parameter.
 In figs 7-9 we show the crossing $-1$ behavior of dark energy. Fig 7 displays the evolution of the
 density of dark energy, figs 8 and 9 illuminate the corresponding EOS and deceleration parameter.

\begin{figure}
\begin{center}
\includegraphics[scale=.6]{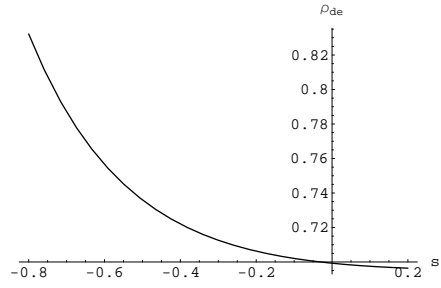}
\caption{The evolution of dark energy density with respect to $s=\ln a$. In this figure, the
parameters are taken as follows: $l = 1, n = 0.7$, ${\Omega}_{m0}
$=0.29, $\widetilde{\Omega}_{m0}=0.3$. We see that the dark energy evolves as quintessence.}
\end{center}
\label{rhodequi}
\end{figure}

\begin{figure}
\begin{center}
\includegraphics[scale=.6]{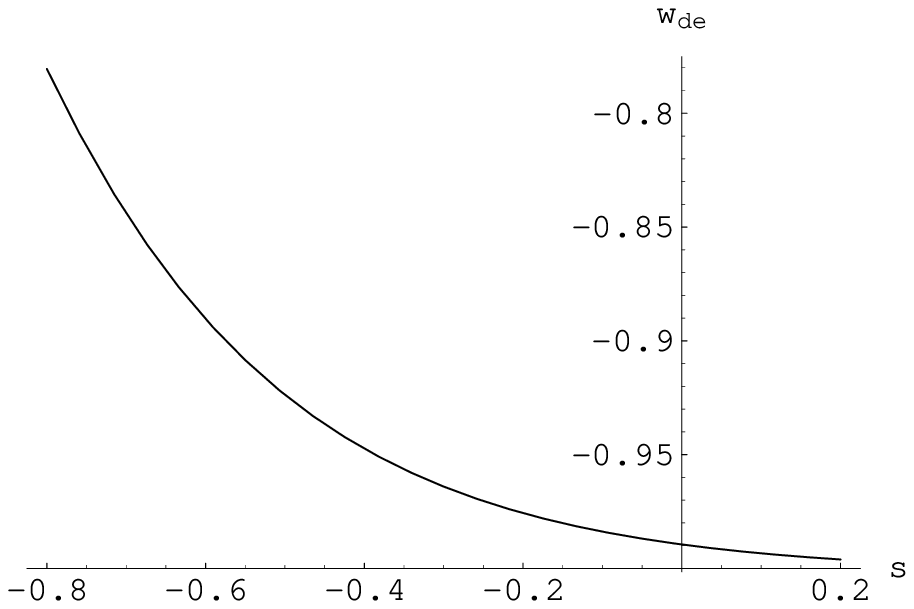}
\caption{$w_{de}$ evolves as a function of $s$. We set the same parameter set as of figure 1. It is
clear that $w_{de}$ is always larger than $-1$.}
\end{center}
\lb{wdequin}
\end{figure}

\begin{figure}
\begin{center}
\includegraphics[scale=.6]{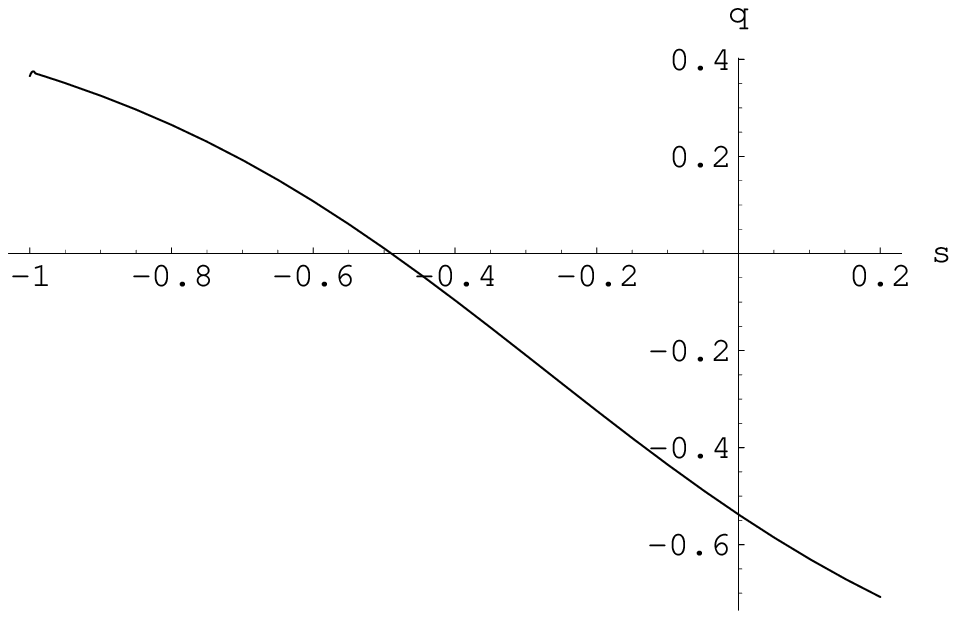}
\caption{$q$ evolves as a function of $s$.  We set the same parameter set as of figure 1. One sees that
$q\sim -0.6$ at present epoch and becomes positive at high redshift, which is consistent with observations.}
\end{center}
\lb{qquin}
\end{figure}

  \begin{figure}
\begin{center}
\includegraphics[scale=.6]{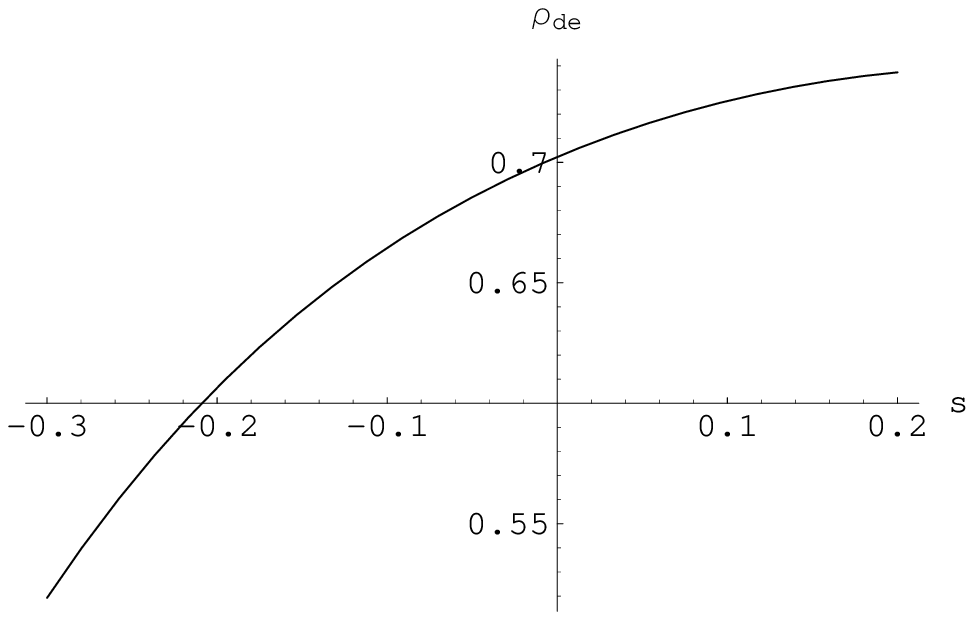}
\caption{The evolution of dark energy density with respect to $s=\ln a$. In this figure, the
parameters are taken as follows: $l = 1, n = 0.8$, ${\Omega}_{m0}
$=0.27, $\widetilde{\Omega}_{m0}=0.3$. We see that the dark energy evolves as phantom.}
\end{center}
\lb{rhodephan}
\end{figure}

 \begin{figure}
\begin{center}
\includegraphics[scale=.6]{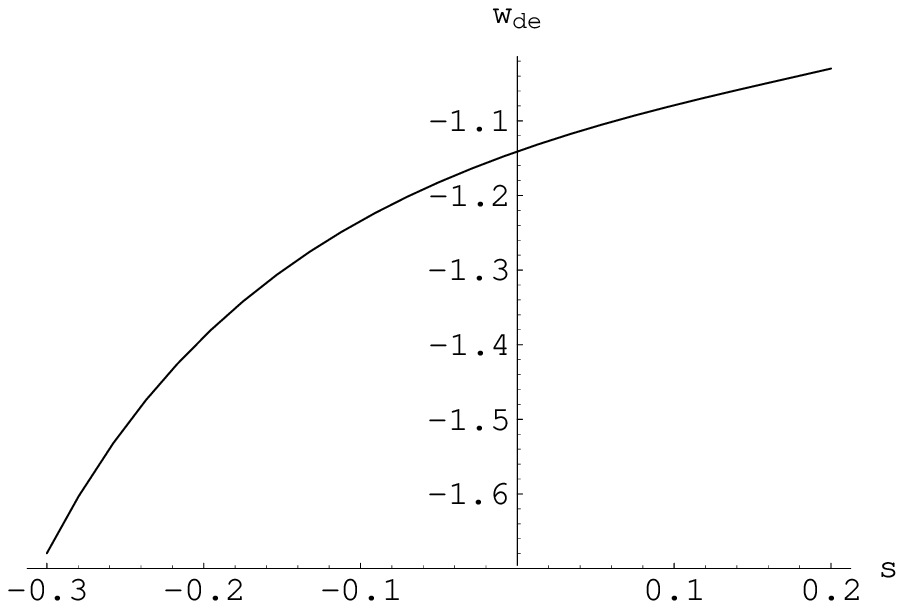}
\caption{$w_{de}$ evolves as a function of $s$. We set the same parameter set as of figure 4. It is
clear that $w_{de}$ is always less than $-1$.}
\end{center}
\lb{wdephan}
\end{figure}

 \begin{figure}
\begin{center}
\includegraphics[scale=.6]{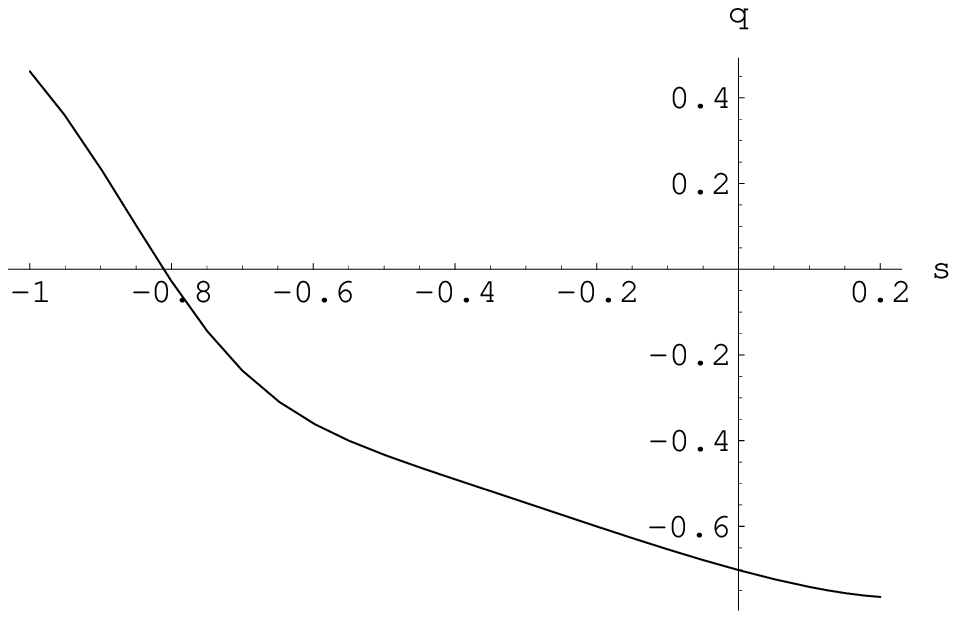}
\caption{$q$ evolves as a function of $s$.  We set the same parameter set as of figure 4. One sees that
$q\sim -0.6$ at present epoch and becomes positive at high redshift, which is consistent with observations.}
\end{center}
\lb{qphan}
\end{figure}

\begin{figure}
\begin{center}
\includegraphics[scale=.6]{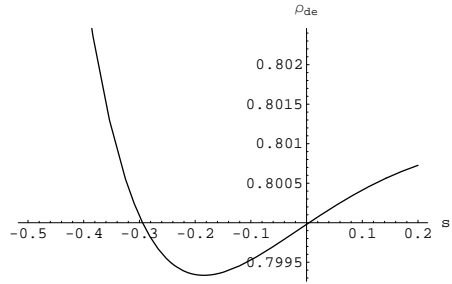}
\caption{The evolution of dark energy density with respect to $s=\ln a$. In this figure, the
parameters are taken as follows: $l = 0, n = 0.8$, ${\Omega}_{m0}
$=0.18, $\widetilde{\Omega}_{m0}=0.2$. We see that the dark energy evolves from quintessence to phantom.}
\end{center}
\lb{rhodecr}
\end{figure}

\begin{figure}
\begin{center}
\includegraphics[scale=.6]{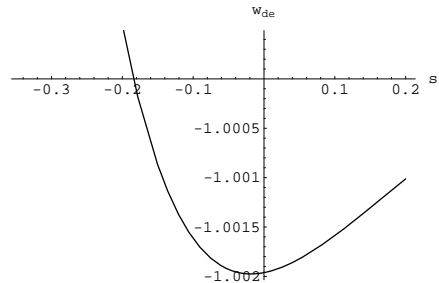}
\caption{$w_{de}$ evolves as a function of $s$. We set the same parameter set as of figure 7. It is
clear that $w_{de}$ crosses phantom divide at about $s=-0.2$.}
\end{center}
\lb{wdecr}
\end{figure}

\begin{figure}
\begin{center}
\includegraphics[scale=.6]{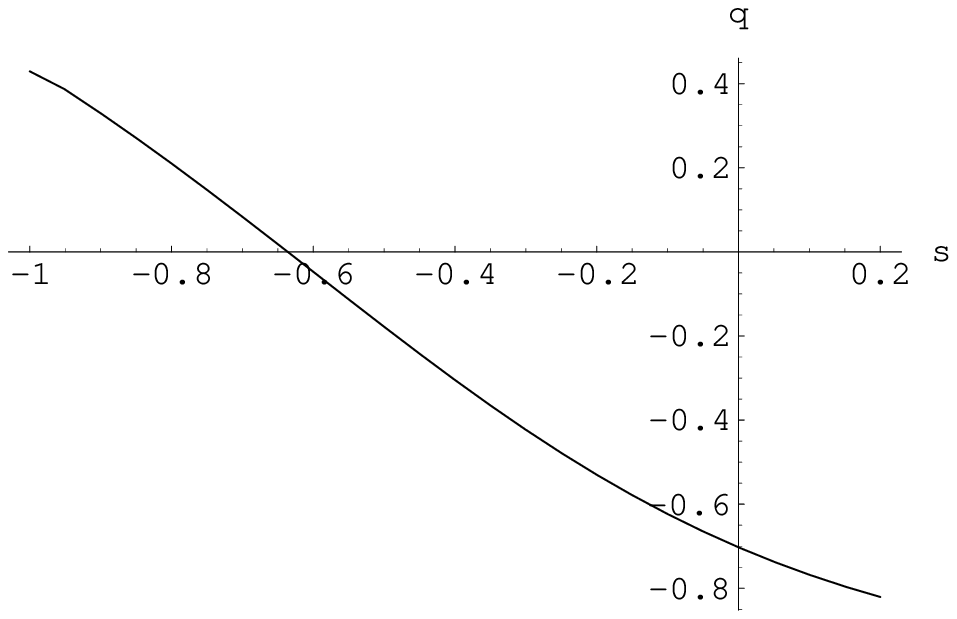}
\caption{$q$ evolves as a function of $s$.  We set the same parameter set as of figure 7. One sees that
$q\sim -0.6$ at present epoch and becomes positive at high redshift, which is consistent with observations.}
\end{center}
\lb{qcr}
\end{figure}

  Note that in figs 7-9, we set $l=0$, which means the potential is a constant. This is a very interesting case in which
  the scalar field rolls on a flat potential, which generates an effective dark energy crossing the phantom divide.

  From figs 1-9, we see that the dark energy in STG has rich properties with a very simple potential (\ref{vnew}).
  Qualitatively, the extra term $(1-\kappa \xi \phi^2)$ plays a critical role to yield the differences from general relativity.
  For example, though $\rho_{de}$ decreases, the variation of $(1-\kappa \xi \phi^2)$ can weaken, counteract, or even turn over this
  trend.

\section{Conclusion and discussion}

 Exact solution is a substantial topic in any physical theories, especially in the non-linear theories, since
 generally the algebraic sum of two independent solutions is not a solution. For a non-linear theory,
 the properties of approximate solution may be far away from the real exact solution.  Every new solution is unique. The
 STG is also highly non-linear as Einstein's theory, even more than it. Though numerous works have been done
 in the area of approximate de Sitter space in STG \cite{noninf}, the exact de Sitter is not found in the previous works.
 Once we have such a solution, our further works in the early universe and in late-time universe will be founded on
 solid rocks, since both the early universe (inflationary stage) and the late-time universe (present cosmic acceleration) are quasi-de Sitter phases.

 We first find a de Sitter solution of STG. In this solution, the effective Newtonian constant is
 a function of time, which exactly counteracts the effects of varies of density and pressure. Thus the spacetime holds
 maximally symmetric.   The potential of the non-minimal
 coupling scalar takes a power-law form. In this solution, the scalar rolls down from a power-law
 potential though the geometry is exactly a de Sitter.

 Based on this solution, we explore cosmology in frame of STG. We find the single
 non-minimal scalar can simulate quintessence, phantom, and can cross the phantom divide.

 As we have mentioned, this exact solution also can be applied to the early universe. In the forthcoming paper, we
 study the inflation based on this solution.

 {\bf Acknowledgments.}
 This work is supported by National Education Foundation of China under grant No. 200931271104 and
 Shanghai Municipal Pujiang grant.

\end{document}